\documentstyle[11pt]{article}
\newcommand{\dfr}[2]{\frac {\displaystyle #1}{\displaystyle #2}}
\begin{document}
\title{Description of Fischer Clusters Formation in Supercooled
Liquids Within Framework of Continual Theory of Defects}
\author{M G Vasin,  V I Lad'yanov}
\maketitle
\begin{abstract}
Liquid is represented as complicated system of disclinations according to defect description of liquids and glasses. The expressions for the linear
disclination field of an arbitrary form and energy of inter-disclination interaction are derived in the framework of gauge theory of defects. It
allows us to describe liquid as a disordered system of topological moments and reduce this model to the Edwards--Anderson model with large-range
interaction. Within the framework of this approach vitrifying is represented as a "hierarchical" phase transition. The suggested model allows us to
explain the process of the Fischer clusters formation and the slow dynamics in supercooled liquids close to the liquid--glass transition point.
\end{abstract}

\begin{section}{Introduction}
The Fischer clusters are long-range correlations of density
fluctuations, which are observed in the supercooled liquids ($\sim
$100 K above the glass transition temperature
$T_g$)~\cite{Fischer}. To date these clusters have been discovered
in the polymers, glassforming, as well as in single fluids. The
typical size of these formations is $\sim $100--300~nm, whereas
the correlation radius of usual short-range thermal density
fluctuations is $\sim $1~nm. The fractal-like structure and long
lifetime are important typical properties of the Fischer
clusters~\cite{Bakai}. There are theoretical works in which the
formation of these clusters is supposed to be the process of
condensation of atom groups with a common type of local atomic
ordering~\cite{Bakai, Tanaka}. However, in the case of dense
liquids (at low temperatures) in these models have complications
in definition of order parameter. Besides this approach does not
allow us to explain the roots of fractal-like structure of these
formations. Therefore the nature of the Fischer clusters is not
quite clear yet.

In the present work the theoretical model of glass transition,
based on the known approaches of the disordered systems physics,
is offered. Our model is based on the disclination model of
amorphous structures, suggested by D. Nelson. From the beginning
we write the system Lagrangian and deriv the expressions for the
linear disclination field and energy of the inter-disclination
elastic interaction (2-th section). After that we define the
topological moment of the disclination system and describe liquid
as a disordered system of interacting topological moments. It
allows us to represent the liquid-glass transition in terms of
spin-glass systems physics, and use the spin-glass theory to
estimation of the size of the long-range correlations in
supercooled liquids (3-th section).
\end{section}

\begin{section}{Disclination description of liquids}

The disclination description of liquids, suggested by D. Nelson,
is the basis of our model. The basic thesis of this one is that
the regular tetrahedron represents the closest and most profitable
local atomic packing. Indeed, it has been demonstrated that the
structure of atomic liquids and glasses has a significant
polytetrahedral character \cite{191}, first due to the success of
the close random packing of hard spheres \cite{192} as a model for
metallic glasses and later due to computer simulations \cite{195}.
In the framework of this polytetrahedral model the liquid
structure is considered to be a tessellation of all space by
tetrahedra with atoms at their vertices. However, it is known,
that Euclidean space cannot be paved only with regular
tetrahedrons. With mathematical rigour it was demonstrated that it
was possible only in the case of a 4-hypersphere space
\cite{Math}. D. Nelson demonstrated that in order to transform a
hypersphere into a flat space, it was necessary to introduce
linear defects (like disclination and dislocation) into the
structure \cite{198}. Thus, the system of linear defects is an
integral element of the polytetrahedral structure, and, owing to
its topological stability, disclination can be considered as a
basic structural element of the liquid structure. Hence, we can
represent the liquid structure as a disclination system.

In order to describe the disclination system, we use gauge theory
of defects \cite{K}. To write down the Lagrangian of the defect
system, let us make use of the standard theory of elasticity. The
simplest Lagrangian, describing the system with elastic
deformations, has the form:
$$
  L_0=\dfr 12\rho _0\partial _4\chi _i\partial
  _4\chi _i-\dfr 18 \left[ \lambda \,u_{\alpha \alpha }u_{\beta \beta
  }+2\mu u_{\alpha \beta }u_{\alpha \beta }\right]\,,
$$
where $\chi _i(\bar r,\,t)$ is the elastic strain field, $\lambda
$ and $\mu $ are the Lame constants, $\rho _0$ is the mass density
(which is considered to be constant for simplicity), and $u_{ij}$
are the relative deformation components:
$$
  u_{ab}=C_{ab}-\delta _{ab}=\partial _a\chi ^i\partial
  _b\chi^i-\delta _{ab}
$$
(Greek letters $\alpha ,\,\beta ,\,\ldots $ are used to denote the
space components set $\{ \alpha \}=\{1,\,2,\,3\}$, and Roman
letters $a,\,b,\,\ldots $ are used to denote the full index set,
including the time component $X^4$ $\{a\}=\{1,\,2,\,3,\,4\}$).

According to the gauge theory of dislocations and disclinations,
the plastic deformation of the matter structure can be considered
as breakdown of the homogeneity of the rotation and translation
(SO(3)$\triangleright $T(3)) groups action. In order to take into
account these homogeneity breakdowns, the compensating fields are
introduced into the Lagrangian ($A_a^{\alpha }$ and $\varphi
^i_b$), and transition from ordinary to covariant derivatives is
carried out:
$$
  \partial _a \chi ^i\to B^i_a=\partial _a\chi ^i+\gamma
^i_{\alpha j}A^{\alpha }_a\chi ^j+\varphi ^i_a,
$$
where $\gamma ^i_{\alpha j}$ are three generating matrixes of the
semisimple group SO(3). After that the Lagrangian $L_0$ is
substituted by the new Lagrangian:
$$
  L=L_0+s_1L_1+s_2L_2,
$$
where $s_1$ and $s_2$ are free parameters of the theory, and the
first term describes elastic properties of the matter:
$$
\displaystyle L_0=  \dfr 12\rho _0B^i_4B^i_4-\dfr 18 \left[
\lambda E_{\alpha \alpha }E_{\beta \beta }+2\mu E_{\alpha \beta
}E_{\alpha \beta }\right]
$$
(where $E_{ab}=B_{ia}B_{ib}-\delta _{ab}$ is the strain tensor).
The second term,
$$
  \displaystyle s_1L_1=-\dfr 12 s_1D^i_{ab}k^{ac}k^{bd}D^i_{cd},
$$
describes the dislocations, the following denotation is used here:
$$
  \displaystyle D^i_{ab}=\partial _a\varphi ^i_b-\partial _b\varphi
  ^i_a+\gamma ^i_{\alpha j}\left( A^{\alpha }_a\varphi ^j_b
  -A^{\alpha }_b\varphi ^j_a+F^{\alpha }_{ab}\chi ^j\right)
  \quad
  (F^{\alpha }_{ab}=\partial _aA^{\alpha }_b-\partial _bA^{\alpha }_a+
  C^{\alpha }_{\beta \gamma }A^{\beta }_aA^{\gamma }_b),
$$
in general case $C^{\alpha }_{\beta \gamma }$ are constants of the
structure of the semisimple group SO(3) ($C^{\alpha }_{\beta
\gamma }=\varepsilon ^{\alpha }_{\beta \gamma }$). The third term,
\begin{equation}\label{1}
  s_2L_2=-\dfr 12s_2C_{\alpha \beta }F^{\alpha
  }_{ab}g^{ac}g^{bd}F^{\beta }_{cd} \qquad
  (g^{\alpha \beta }=-\delta ^{\alpha \beta },\quad
  g^{44}=1\left/\zeta\right. ),
\end{equation}
describes disclintions. The Yang-Mills fields, $A^{\alpha }_a$,
and $\varphi ^i_a$ describe disclinations and dislocations
accordingly.

Usually it is supposed that the disclinations are most important
structural elements in the description of amorphous matter, since
they have the largest energy density $s_2\gg s_1$, and determine
two fundamental properties of glasses: the absence of long-range
ordering and the resistance to crystallization. It is natural to
suppose that disclinations are also fundamental structure elements
in liquids, therefore hereinafter we will focus only on the case
of breakdown of homogeneity of the rotation semi-group (monoid).
In other words, we will neglect the $\varphi ^i_a$ field
contribution to the action and restrict ourselves to the
consideration of the theory with purely disclination Lagrangian,
$$
  L=L_0-s_2L_2.
$$

In the case of linear disclination the difficulties, concerning
nonlinearity of the Yang-Mills fields, can be avoided. The field
of the disclination line element differs from the point
disclination field, since in this case the rotational
displacements, corresponding to the gauge transforms, can be
performed only around the tangent to the defect line vector,
directed along this element (wedge disclination), and,
consequently, the gauge group reduces to SO(2). Therefore, the
theory becomes much simpler since the rotation group SO(2) is
Abelian instead of the non-Abelian group SO(3) (In work
\cite{Osipov} this fact was used in the case of infinite
rectilinear defects). In order to find the field of linear
disclination, let us introduce the tensorial fields $G_{k}^{\alpha
}$ and $P_{k}^{\alpha }$, which take the following form:
$$
\displaystyle G_{k}^{\alpha }\equiv F_{k4}^{\alpha }=\partial
_kA_4^{\alpha },\qquad \displaystyle  P_{k}^{\alpha }\equiv \dfr
12 \varepsilon _{klm}F_{lm}^{\alpha }=\dfr 12 \varepsilon
_{klm}\left[ \partial _lA_m^{\alpha }-\partial _mA_l^{\alpha
}\right],
$$
then the energy functional of the ''free'' disclination field
(\ref{1}) can take the form of:
$$
L_2=\dfr 14\int (d^3x)F^{\alpha }_{ij}F^{\alpha }_{ij}=\dfr 12\int
(d^3x)\left[ (P^{\alpha }_k-G^{\alpha }_k)(P^{\alpha }_k-G^{\alpha
}_k)\right]+\int (d^3x)P^{\alpha }_kG^{\alpha }_k.
$$
Minimizing this part of the Lagrangian we can get the
quasi-stationary expression for the potential of the gauge field
of the wedge disclination fragmenton $dl_{\alpha }$:
$$
A_4^{\alpha }=x_{\alpha }\left( \dfr C{r^3}+2C_1\right) dl_{\alpha
}, \qquad A_k^{\alpha }= \, \varepsilon _{\alpha kj}x_j\left( \dfr
C{r^3}+C_1 \right)dl_{\alpha },
$$
where $C$ and $C_1$ are arbitrary constants. It should be noted
that in this expression there is no summing with respect to
$\alpha $, and $C_1=0$ because of $A_k^{\alpha }(\infty )=0$.

Let us consider the two basic examples: 1) First we find the field
of the infinite rectilinear disclination, directed along the axis
$i_{\alpha }$. For that we introduce the normal to the axis
$i_{\alpha }$ vector $R_j$, which determines the point location
with respect to disclination. Then the expression of the field of
the rectilinear disclination, directed along the axis $z$, can be
written in the form of:
$$
A_{i}^z=C \, \varepsilon _{zij}\int\limits_{-\infty }^{\infty}
\dfr {x_j}{r^3} dl_{z}=C \, \varepsilon _{\alpha
ij}\int\limits_{-\infty }^{\infty} \dfr {R_j}{\left(
R^2+{l_{z}}^2\right) ^{3/2}}dl_{z}=2C \, \varepsilon _{z ij}\dfr
{R_j}{R^2}.
$$
Due to the condition $\Omega = 2\pi \nu=\oint A^z_idl_i$, where
$\Omega $ is the Frank vector, and $\nu $ is the Frank index, we
get the expression which agrees with the result of \cite{Osipov}.
2) The field of the element of the round (with the radius equal to
$a$) disclination loop has the form of:
$$
A_{i}^{\alpha }=\dfr {\nu }2\,\varepsilon _{\alpha ij}\dfr
{x_j}{r^3} dl_{\alpha }= \dfr {\nu a}2 n_i\dfr 1{r^3} dl_{\alpha
},
$$
where $n_i$ is the unit vector normal to the loop's plane.

If we consider only the low-angle disclinations, $\nu \sim 0.2$,
which are typical for the liquid-like structure, the stress field
around the disclination line has the form of:
$$
\sigma _{ai}=\mu E_{ai}=\mu \left( B_{aj}B_{ji}-\delta
_{ai}\right)=\mu \left[ \varepsilon _{\alpha al}x_lA^{\alpha
}_i+\varepsilon _{\alpha il}x_lA^{\alpha }_a+\partial _i
u_a+\partial _au_i\right] .
$$
Hence, the volumetric density of elastic interaction energy of
disclnations, described by the fields $A$ and $A'$, has the form
of:
$$
F= \dfr 12 \sigma _{ai} E_{ai} \simeq \dfr {\mu }2\varepsilon
_{\alpha al}x_lA^{\alpha }_i \varepsilon _{\gamma
aj}x_j'A'^{\gamma }_i +O(a^4).
$$
One can see that form of this interaction is similar to the form
of electrodynamic interaction of currents. For example, the
elastic interaction energy of two round loops $\Gamma $ and
$\Gamma '$ with the radius $a$, being remote from $\vec r$, has
the form:
\begin{equation}\label{EI}
F\simeq n_in_i'\dfr {\pi ^2\mu a^2\nu \nu '}4\oint\limits_{\Gamma
} \oint\limits_{\Gamma ' } \dfr {1}{r}\,dl_jdl_j'.
\end{equation}
This expression agrees with the expression of the interaction
energy of current loops on condition that $\vec n\|\vec n'$.
\end{section}

\begin{section}{Description of glass transition}
The basic idea of our approach was first suggested by N. Rivier
\cite{Riv}. According to his theory the glass transition can be
considered as phase transition in the system of topological
defects. In this theory there is no principal distinction between
the spin or structure-disordered systems. Both in the first and in
the second cases the problem of description of the system is
reduced to the description of the system of interacting
disclinations. We would like to make use of this analogy and
reduce this description to that of the disordered spin systems as
much as possible, and apply the methods, known in the spin glass
physics, to our system.

Computer simulation results testify that the disclination system
is a tangled net of interacting disclinations. To describe this
system, let us make use of the approach, which is well-known in
classical electrodynamics \cite{Ioffe}: we associate the system of
linear disclinations, included in some small volume, such that the
numbers of positive and negative disclinations entering this
volume are equal, with their general topological moment (the
analogue to the magnetic moment of the electric currents system).
Thus, the total elastic interaction energy between all the
elements of the disclination network can be represented as the
interaction energy of the system of the local topological moments.
Since the disclination loop is the simplest structure element,
having a topological moment, to simplify the model, let us imagine
the disclination network as the system of randomly located and
randomly directed disclination loops.

Let us determine the topology moment vector $\bar S_i$ in our
system as magnetic moment is determined in electrodynamics. In
this case the Hamiltonian, that describes the disclination loops
system, can be represented in the form of Heisenberg model
Hamiltonian:
$$
   \displaystyle F=\dfr 12\sum\limits_{i,\,j}J_{ij}\vec S_i \vec S_j, \\
$$
with the inter-spin coupling (from (\ref{EI}))
$$
   J^{lk}_{ij}\approx {\mu \pi ^4 a^6 \nu _i\nu _j}\dfr {\cos
   (\angle \vec S_i\vec S_j)}{2|\vec r_{ij}|^3}{\left(
   \delta ^{lk}-3\dfr {r^{l}_{ij}r^{k}_{ij}}{r^2_{ij}}\right)} ,
$$
where $l$ and $k$ are the indexes of the space coordinates of the
vector $\bar r_{ij}$. The module and sign of the disclination
loops coupling depend both on their situation and their mutual
orientation. This system is frustrated. Indeed, from the last
expression one can see that the moment coupling is an alternating
quantity, hence, with equal possibility these moments can interact
in either ferromagnetic-like or antiferromagnetic-like manner
($\langle J_{ij}\rangle =0$). This character of coupling is
typical for spin glasses~\cite{Dot}. Therefore, we attempt to
describe the considered system, using the modification of the
Edwards-Anderson model with large but finite-range interaction,
which was analyzed in \cite{Dot, IF}. According to this analysis,
glass transition in such systems is the so-called ''hierarchical''
phase transition, which is hierarchy of successive transitions
that closes at the glass transition point $T_g$. Using the results
of \cite{IF} and the expression of defect coupling energy obtained
above one can estimate the transition points $T_i$ for every
$i$-th step of this series, as well as the correlation lengths
corresponding to these points: $T_i$ can be estimated from:
\begin{equation}\label{15}
 \displaystyle \ln \left[ \frac{\displaystyle T_i-T_g}{\displaystyle
 T_g}\right] \approx \displaystyle \left[-\frac p{q-1}(q^{1+i}-1)\right] \ln Z_0 ,
\end{equation}
where $q=32/15$, $p=4/15$ \cite{IF}, the coordination number,
$Z_0$, can be evaluated from the coupling radius $R$:
$$
  \displaystyle Z_0=\dfr 43c\pi R^3,
$$
where $c$ is disclination density, and $R$ can by determined from
the inter-disclination interaction energy and thermal fluctuations
energy parity condition:
\begin{equation}\label{16.5}
E(A)=\frac {\displaystyle \mu a^6\pi ^4\nu ^2}{\displaystyle
2R^3}\approx \dfr {3kT}2,
\end{equation}
where $a$ is a typical interatomic distance. If we express $R$
from (\ref{16.5}) and substitute this expression in (\ref{15}) we
find that the freezing process begins at the temperature, which
can be estimated from solving the equation
\begin{equation}\label{17}
  \displaystyle T_0\approx T_g\left[ 1+Z_0^{-\frac
  4{15}}\right]=T_g\left[ 1+\left[\dfr {4\mu c\pi^5a^6\nu ^2}{9kT_0}\right] ^{-\frac
  4{15}}\right].
\end{equation}
Subsequent cooling down results in formation of the clusters which
finally have shaped at the temperature:
$$
\displaystyle T_1\approx T_g\left[ 1+Z_0^{-0.84}\right].
$$
The correlation length characterizing the average size of these
clusters as well as their average life time can be determined as:
\begin{equation}\label{18}
  \displaystyle \xi \approx\left( \frac {Z_0^{2q}}{4 c}\right) ^{\frac 13},
  \qquad \tau \approx\tau _0Z_0^{14/5}.
\end{equation}
For quantitative estimation of these parameters let us assume
their typical values: $a=1\cdot 10^{-9}$~m; $k=1.38\cdot 10^{-23}$
J/K, $T_g=200$~K; $\tau _0\approx 10^{-10}$ s; the estimation of
the disclination density follows from the analysis of the
experimental data (typical interdisclination distance is $\sim
5a$): $c=1.4\cdot 10^{24}$ m$^{-3}$; and the shear modulus of
liquids is $\mu \approx 10^{10}$ N/m$^2$, according to
\cite{Frencel}. As a result, after substitution of these
parameters in (\ref{17}) and (\ref{18}), we get the following
estimations: $T_0-T_g\approx 100$ K; $T_1-T_g\approx 20$ K; $\xi
\approx 300$ nm; $\tau \sim 0.4\cdot 10^{-6}$ s. Of course, these
estimations are rather proximate yet. But we hope, that they can
be improved in the future.
\end{section}

\begin{section}{Conclusions}
The structure frustrations and complicated cooperative character
of atoms moving in liquids lead to the necessity of using of the
state-of-the-art methods of the statistical physics to understand
the nature of its properties. In this paper we have tried to
combine the gauge fields theory and Anderson localization theory
to describe the structure formation processes in supercooled
liquids.

The given above estimations allow us to conclude, that the process
of the freedom degree freezing begins at the temperatures that are
significantly above the glass transition temperature, $T_g$, and
in the temperature range of $(T-T_g)$$\sim $100 K in the structure
of the liquid the correlations with the size $\sim $100 nm form.
The disclination model allows us to explain the density beating by
natural way, since the atom density is larger close to a negative
($\nu >0$) disclination core and smaller close to a positive ($\nu
<0$) one. Therefore, we think that these correlations conform to
the long-range density correlations (Fischer clusters), which are
observed in the experiments.
\end{section}

This study was supported by the RFBR grant (04-03-96020-r2004ural), and grant of FSP ''Integracia'' (B0086/2121).

%\section*{References}


\begin{thebibliography}{99}

\bibitem{Fischer} E.W. Fischer, Laight scattering and dielectric
studies on glass forming liquids, 1993 {\it Physica A} {\bf 201}
183;

\bibitem{Bakai} A.S. Bakai, Long-range density fluctuations in the glass-forming
liquids, 2002 {\it J. Non-Cryst. Solids.}  {\bf 307--310} 623;

\bibitem{Tanaka} H.~Tanaka, General view of a liquid--liquid phase
transition, 2000 {\it Phys. Rev. E }{\bf 62} 6968;

\bibitem{191} D.R. Nelson and F. Spaepen, Polytetrahedral order in condensed
matter, 1989 {\it Solid State Phys.} {\bf 42} 1;


\bibitem{192} J.D. Bernal, Geometry of the structure of monatomic
liquids, 1960 {\it Nature} {\bf 185} 68;

\bibitem{195} H. Jonsson and H.C. Andersen, Icosahedral Ordering in the
Lennard-Jones Liquid and Glass, 1988 {\it Phys. Rev. Lett.} {\bf 60} 2295;

\bibitem{Math} H.S.M. Coxeter, Regular Polytopes, Dover, New York,
1973;

\bibitem{198} D.R. Nelson, Liquids and Glasses in Spaces of Incommensurate
Curvature, 1983 {\it Phys. Rev. Lett.} {\bf 50}, 982;

\bibitem{K} A.Kadich and D.G.B.Edelen, A
gauge theory of dislocations and disclinations, Springer-Verlag,
Berlin-Heidelberg-New York, 1983;

\bibitem{Osipov} V.A. Osipov, Aharonov-Bohm effect in planar systems
with disclination vortices, 1992 {\it Phys. Lett. A} {\bf 164}
327;

\bibitem{Riv} N. Rivier, D.M. Duffy Line defects and the glass
transition. --- in: Numerical methods in the study of critical
phenomena. Proc. Colloq., Carry-le Rouet, France, June 2--4, 1980/
Eds. J. Della Dora, J. Demongeot, B. Lacolle.--Berlin, Heidelberg,
New York: Springer-Verlag, 1981, pp. 132--142;

\bibitem{Ioffe} I.E. Tamm, Foundation of the electricity (Rus.), Moscow,
Science, 1989, 504. - ISBN 5-02-014244-1;

\bibitem{Dot} V.S. Dotsenco, M.V. Feigel'man and L.B. Ioffe Spin
glasses and related problems, 1990 {\it Sov. Sci. Rev. A. Phys.}
{\bf 15} 1--250;

\bibitem{IF} L.B. Ioffe and M.V. Feigel'man, Hierarchical structure
of the Edwards-Anderson spin glass, 1985 {\it Sov.Phys.\,JETP}
{\bf 62} 376;

\bibitem{Frencel} Y.I. Frencel, Kinetic theory of liquid (Rus.), AS USSR, Moscow, 1945, 424 p.

\end{thebibliography}
\end{document}